\documentclass[referee]{cjaa}           

\usepackage{amsmath}
\usepackage{graphicx}                   
\input{epsf.sty}                        
\input{psfig.sty}                       

\begin{document}

   \title{Statistical Analysis of Point-like Sources in {\it Chandra} Galactic Center Survey}

   \volnopage{Vol.0 (200x) No.0, 000--000}      
   \setcounter{page}{1}          

   \author{J. F. Wu
      \inst{1}\mailto{}
   \and S. N. Zhang
      \inst{1,2}
   \and F. J. Lu
      \inst{2}
   \and Y. K. Jin
      \inst{3}
      }
   \offprints{J. F. Wu}                   

   \institute{Department of Physics and Center for
Astrophysics, Tsinghua University, Beijing 100084, P. R. China\\
             \email{jfwu03@mails.tsinghua.edu.cn}
        \and
             Key Laboratory of Particle
Astrophysics, Institute of High Energy Physics, Chinese Academy of
Sciences, Beijing 100049, P. R. China\\
        \and
             Department of Engineering Physics and Center for
Astrophysics, Tsinghua University, Beijing 100084, P. R. China\\
          }

   \date{Received~~2006 month day; accepted~~2006~~month day}

   \abstract{
{\it Chandra} Galactic Center Survey detected $\sim 800$ X-ray
point-like sources in the $2^{\circ} \times 0.8^{\circ}$ sky
region around the Galactic Center. In this paper, we study the
spatial and luminosity distributions of these sources according to
their spectral properties. Fourteen bright sources detected are
used to fit jointly an absorbed power-law model, from which the
power-law photon index is determined to be $\sim$2.5. Assuming
that all other sources have the same power-law form, the relation
between hardness ratio and HI column density $N_H$ is used to
estimate the $N_H$ values for all sources. Monte Carlo simulations
show that these sources are more likely concentrated in the
Galactic center region, rather than distributed throughout the
Galactic disk. We also find that the luminosities of the sources
are positively correlated with their HI column densities, i.e. a
more luminous source has a higher HI column density. From this
relation, we suggest that the X-ray luminosity comes from the
interaction between an isolated old neutron star and interstellar
medium (mainly dense molecular clouds). Using the standard Bondi
accretion theory and the statistical information of molecular
clouds in the Galactic center, we confirm this positive
correlation and calculate the luminosity range in this scenario,
which is consistent with the observation ($10^{32}\sim 10^{35}$
ergs s$^{-1}$).
   }

   \authorrunning{J. F. Wu, S. N. Zhang, F. J. Lu \& Y. K. Jin }            
   \titlerunning{Statistical Analysis of Point-like sources in {\it Chandra} GCS}  

   \maketitle

%
%
\section{Introduction}           
\label{sect:intro}
The central regions of galaxies are usually crowded by many
celestial bodies with different physical properties. The center of
our Galaxy is an ideal laboratory to study X-ray sources and the
related high energy astrophysical processes. It has been the
observational target of most X-ray satellites, such as {\it ROSAT}
(Snowden {\it et al.} 1997; Sidoli, Belloni \& Meregetti 2001),
{\it ASCA} (Sakano {\it et al.} 2002) and {\it BeppoSAX} (Sidoli
{\it et al.} 1999). In these past observations, many X-ray
point-like sources have been detected. The {\it ROSAT} survey
covered a $3^{\circ} \times 4^{\circ}$ sky region around the
Galactic Center and detected 107 X-ray point-like sources in
0.1-2.4 keV. In {\it ASCA} observations, 52 point-like sources
were detected. {\it BeppoSAX} has also observed 16 X-ray
point-like sources in the central region of our Galaxy. {\it
Chandra X-ray Observatory} ({\it CXO}) has much better spatial
resolution (0.5 arcsec) than previous satellites (e.g. {\it
ROSAT}\ 20 arcsec). It can produce high quality images of the
Galactic center. {\it Chandra} Galactic Center Survey (hereafter
GCS) was carried out in July, 2001 (Wang, Gotthelf \& Lang 2002).
It consists of 30 separate ACIS-I observations (Obs. ID 2267-2296,
energy band 0.2-10.0 keV) with a total exposure time of 94.2 hours
($\sim 340$ ks). The sky area covered is $2^{\circ} \times
0.8^{\circ}$, about 280 pc $\times \ 110$ pc (taking the distance
of Galactic center to be 8.0 kpc). About 800 point-like sources
were detected.

A variety of analyses to explore the nature of these discrete
sources have been done. Pfahl, Rappaport \& Podsiadlowski (2002)
discussed the possibility of wind-accreting neutron star nature of
$Chandra$ GCS sources, while Belczynski \& Taam (2004) introduced
another possibility for low luminosity ($10^{31}\sim 10^{32}$ ergs
s$^{-1}$) source nature, the Roche lobe overflow accretion
systems. Belczynski \& Taam (2004) also argued that accreting
neutron star systems are not likely to be the nature of the major
part of the sources. Multiwavelength observations have been
proposed on these sources to further explore the nature of these
sources. ChaMPlane ({\it Chandra} Multiwavelength Plane Survey,
Grindlay {\it et al.} 2005) focuses on the multiwavelength
observations on the $Chandra$ X-ray sources in the Galactic plane
and bulge. They employed X-ray and optical surveys, and a followup
spectroscopy and an IR identification program. Their recent
results (Laycock et al. 2005) showed that high mass X-ray binaries
(HXMBs) are not the dominant population (fewer than 10\%) of the
X-ray sources in the Galactic center. Bandyopadhyay {\it et al.}
(2005) took an infrared survey on parts of the $Chandra$ GCS
region. Analysis on potential counterparts indicated that the
sources may be accreting X-ray binaries. Recently, Liu \& Li
(2005) carried out an evolutionary population synthesis study on
discrete sources in the Galactic center region. Their results
showed neutron star low mass X-ray binaries could not account for
most sources in Wang et al. (2002) survey. Muno et al. (2006)
reanalysed the data of {\it Chandra} GCS and suggested that major
part of these sources could be CVs since the local Galactic
neighborhood X-ray sources showed a CV population in the $10^{32}
- 10^{34} \ {\rm ergs \ s^{-1}}$ luminosity range.

Despite of extensive work done previously, the nature of X-ray
sources in the Galactic center is still not understood completely.
Our work presented here is a statistical analysis of the sources
in Wang's $Chandra$ GCS, including their spatial distribution and
flux characteristics. In \S 2, data analysis method and results
are described in detail, including spectral fitting and parameter
determination. In \S3, the spatial distribution of the sources is
discussed. In \S 4, another potential nature of GCS X-ray sources
is proposed. \S 5 is a summary.

\begin{figure}[tb]
 \begin{minipage}[t]{0.48\linewidth}
  \centering
  \includegraphics[scale=0.35]{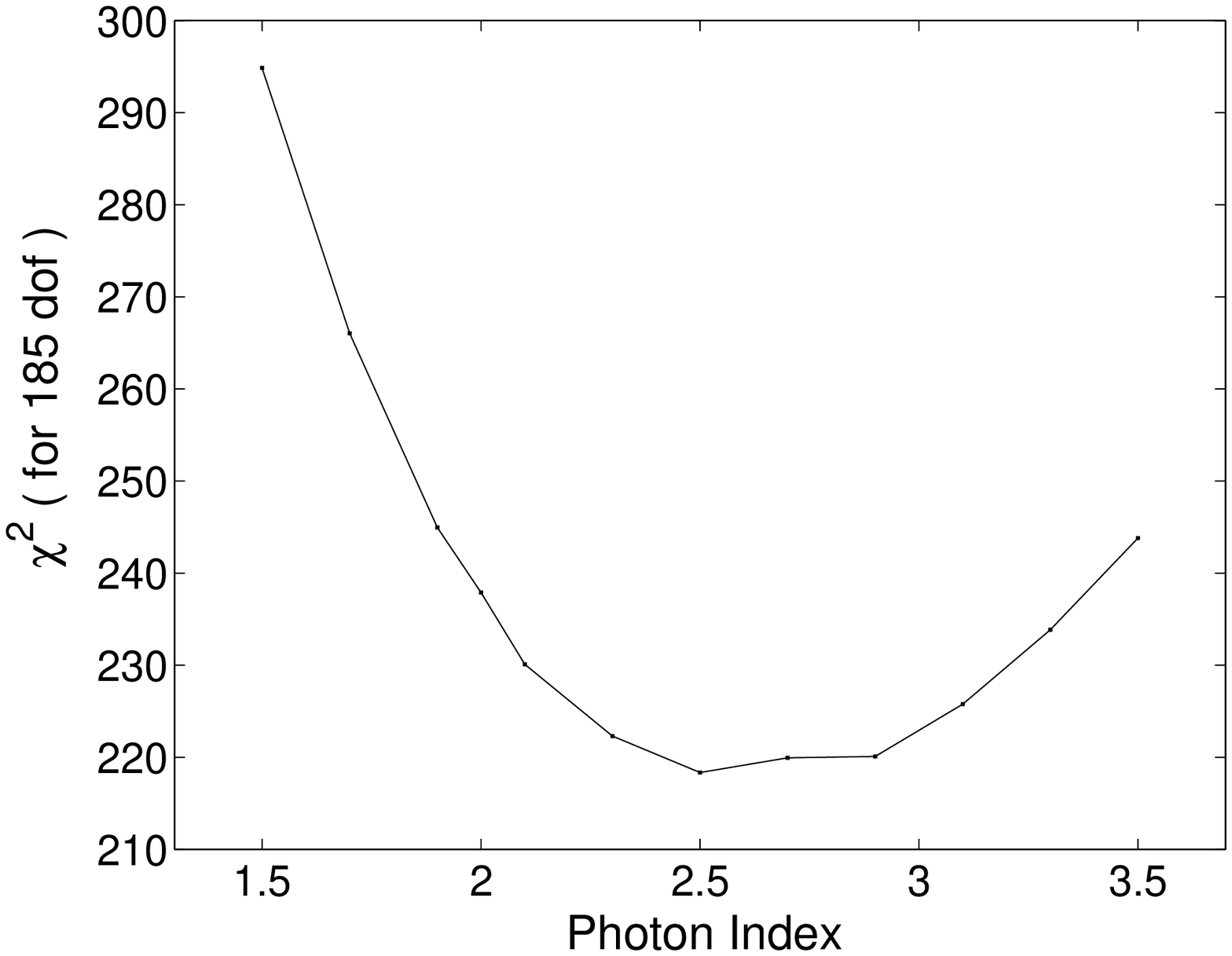}
  \vspace{-0.2in}
  \caption[]{The total chi-squared values for different photon indices.
  When $\Gamma$=2.5, the reduced chi-squared value is most close to 1.}
  \label{fig:side:a}
 \end{minipage}\ \ \ \ \ \ \
 \begin{minipage}[t]{0.48\linewidth}
  \centering
  \includegraphics[origin=c,angle=270,width=2.5in,height=1.9in]{fig2.ps}
  \vspace{-0.2in}
  \caption[]{Fitted spectrum of CXO
  J174417.2-293945:
  $N_H=(0.5476\pm0.0876) \times 10^{22}\ {\rm cm}^{-2}$, $\Gamma$=2.5.}
  \label{fig:side:b}
 \end{minipage}
\end{figure}

\bigskip


\section{Spectral fitting \& $N_H$ determination}
\label{sect:Spec}
{\it Chandra} has excellent angular resolution. Although many GCS
sources have only a few counts, the background counts are even
much lower. So for most sources, the detection probability is
close to unity. The efficiency correction is ignored here because
the luminosity function of these sources is not discussed in this
paper.

{\it Psextract} procedure in the CIAO package is used to get the
spectral files of the sources. The regions we used to extract
energy spectra are circles chosen to contain all source photons,
while minimizing the effect of nearby sources and background. The
radii of these circles range from 6 pixels to 24 pixels. The
background region is made of several circles of similar radii for
these source. The spectra are re-binned to make each energy
channel have no less than 25 counts for most sources. For other
sources with smaller counts, each energy channel has no less than
10 counts. Then the {\it Xspec} in Heasoft package is used to fit
the spectra with the absorbed power-law model, i.e. ``phabs(po)"
model in {\it Xspec}. There are three parameters in this model
--- the photon spectral index $\Gamma$, the neutral hydrogen column density $N_H$ and the
normalization factor $A$. Because of the short effective exposure
time for each field, the total counts for sources without
significant pileup are not sufficient for a high quality spectral
fitting, in order to determine $N_H$ and $\Gamma$ simultaneously
for each source; many other sources are also too faint for any
meaningful spectral fitting individually. We thus choose 14
sources, each with total counts of more than 80 and without
significant pileup (the maxim pileup percentage of the 14 sources
is 8\%). We assume that all sources have a similar value of
$\Gamma$, thus attributing the different hardness ratios of these
sources to different values of $N_H$. Our goal is to search for
one value of $\Gamma$ which best describes all 14 sources.
Spectral fittings of these 14 sources show their power-law photon
indices $\sim 2-3$, indicating soft spectral properties. $\Gamma$
values are tried from 1.5 to 3.5 with step of 0.2. For each trial
value of $\Gamma$, i.e., we fit the absorbed power-law model by
fixing $\Gamma$, the fitting generates one chi-squared value. Then
the total chi-squared values of the 14 sources for all different
values of $\Gamma$ are calculated. We find that $\Gamma=2.5$
corresponds to the minimum chi-squared value (Fig. 1), with a
reduced chi-squared value of about 1.180 for 185 degrees of
freedom. We thus conclude that $\Gamma=2.5^{+0.9}_{-0.5}$ for a
68.3\% confidence interval. We believe this single photon index
can describe most of the GCS sources adequately and no obvious
selection effects exist, because these 14 sources distribute
through the whole ranges of $N_H$ distribution acquired below
(Fig. 4). Table 1 shows the spectral fitting results for these 14
sources with the photon index frozen at 2.5.

\begin{table}[htb]
\caption{Fitting results of 14 {\it Chandra} GCS sources ($\Gamma
=2.5$ )}
\begin{tabular}{r|c|c|c|c}
\hline Obs.ID & Source & Counts & $N_{H}(10^{22} {\rm cm}^{-2})$ &
$\chi ^2$/dof\\
\hline\hline\ 2271 & J174722.9-280904 & 108 & $16.81 \pm 1.101$ &
16.926/21
\\2273 & J174639.1-285351 & 179 & $(1.464 \pm 0.378)\times 10^{-4}$ & 21.616/14
\\2273 & J174622.7-285218 & 119 & $15.02 \pm 2.732$ & 18.401/11
\\2274 & J174705.4-280859 & 141 & $0.2518 \pm 0.2823$ & 8.073/12
\\2275 & J174319.4-291359 & 133 & $1.234 \pm 0.2266$ & 10.669/11
\\2276 & J174550.4-284921 & 180 & $5.534 \pm 0.641$ & 29.323/19
\\2276 & J174550.4-284911 & 86 & $4.989 \pm 1.992$ & 6.515/5
\\2277 & J174804.9-282917 & 94 & $14.92 \pm 3.407$ & 4.583/6
\\2277 & J174729.0-283516 & 204 & $0.8724 \pm 0.2423$ & 23.174/14
\\2278 & J174417.3-293944 & 695 & $0.7984 \pm 0.0907$ & 26.578/24
\\2282 & J174602.2-291039 & 174 & $1.759 \pm 0.252$ & 18.668/16
\\2285 & J174626.1-282530 & 158 & $4.640 \pm 0.616$ & 13.99/13
\\2294 & J174502.8-282505 & 107 & $0.621 \pm 0.250$ & 12.533/11
\\2295 & J174451.7-285309 & 104 & $3.703 \pm 0.655$ & 6.66/8
\\\hline
\end{tabular}
\end{table}

\begin{figure}[htb]
\begin{minipage}[t]{0.5\linewidth}
\centerline{\epsfxsize 73mm \epsffile{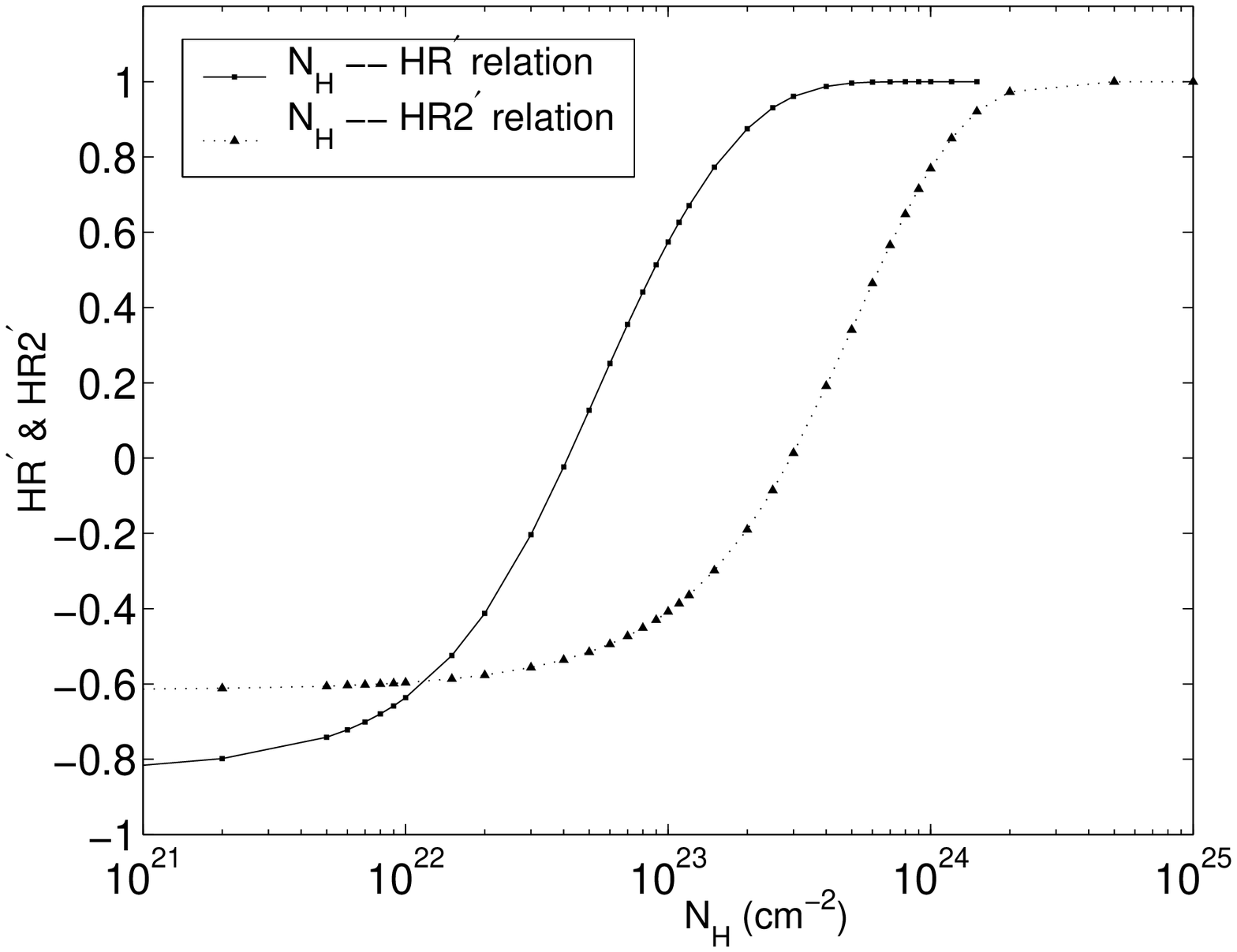}}
\caption[]{Relation between $N_H$ and $HR'$ \& $HR2'$. The solid
line with squares is the relation of $N_H$ and $HR'$, while the
dotted line with triangles is the one of $N_H$ and $HR2'$.}
\label{fig3}
\end{minipage}\ \ \ \ \ \ \
 \begin{minipage}[t]{0.5\linewidth}
 \centerline{\epsfxsize 70mm \epsffile{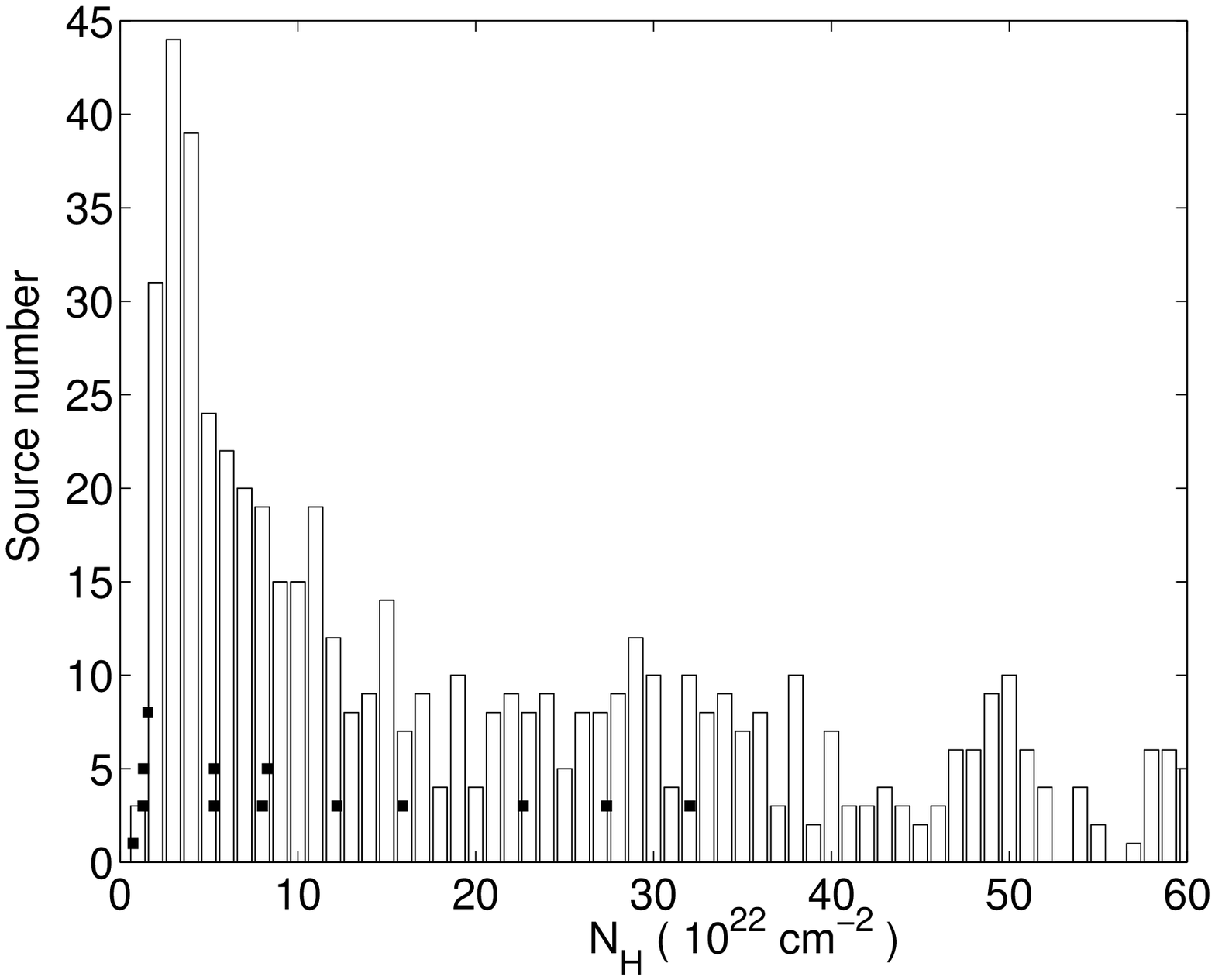}}
\caption[]{The $N_H$ distribution of {\it Chandra} GCS sources.
The black squares are the 13 sources used for determining their
joint photon index (\S 2, one of the 14 sources cannot get the
$N_H$ value using our method); their vertical locations have no
meaning. Since their $N_H$ values span across almost the whole
range, it is believed that there is no selection effect involved.
} \label{fig4}
\end{minipage}
\end{figure}

Assuming that $\Gamma=2.5$ for all sources, $N_H$ value for each
source can be estimated according to the hardness ratios of the
source. Following the definitions of Wang (private communication):
$ HR = \frac{b-a}{b+a}, \ HR2 = \frac{c-b}{c+b}$, where $a,\ b$
and $c$ are counts \footnote{An off-angle efficiency correction
has been applied. However the maximum correction factor for the
8.638 keV photons at 8' off-axis angle is only 1.67. Therefore the
corrected counts still follow Poisson distribution closely.} in
three bands: $A$ (1-3 keV), $B$ (3-5 keV) and $C$ (5-8 keV),
respectively. The values of $HR$ and $HR2$ for each source have
been given in the source list. Using the PIMMS software
(http://cxc.harvard.edu/toolkit/pimms.jsp), the relation between
HI column density $N_H$ and hardness ratios can be obtained as
shown in Fig. 3. Here we should emphasize that the definitions of
hardness ratios in Fig. 3, labelled as $HR'$ and $HR2'$ there, are
different from the ones above, $HR$ and $HR2$. This is because
when using PIMMS, we can only get the values of count rates, i.e.
the parameters of Poisson processes in three bands $\lambda_A,\
\lambda_B,\ \lambda_C$: Their definitions are
$HR'=\frac{\lambda_B-\lambda_A}{\lambda_B+\lambda_A}$,
$HR2'=\frac{\lambda_C-\lambda_B}{\lambda_C+\lambda_B}$.

We therefore use the method in Jin {\it et al.} (2006) to estimate
the values and error intervals of $HR'$ and $HR2'$. For a Poisson
process, the $\lambda$ parameter obeys the Gamma distribution
under certain counts as follows,
\begin{eqnarray}
p(\lambda_A=x|n_A=a)&=&\frac{x^ae^{-x}}{a!}.
\end{eqnarray}
Based on this result, the probability density function of
$\lambda_A/\lambda_B$ is,
\begin{eqnarray}
p(\frac{\lambda_A}{\lambda_B}=z|n_A=a,
n_B=b)&=&\frac{z^a(a+b+1)!}{(z+1)^{a+b+2}a!b!}.
\end{eqnarray}
The expectation of $\lambda_A/\lambda_B$ is
\begin{eqnarray}
E(\frac{\lambda_A}{\lambda_B}|n_A=a, n_B=b) &=& \frac{a+1}{b}.
\end{eqnarray}
The most probable value $z_0$ can be obtained as,
\begin{eqnarray}{z_0}&=&\frac{a}{b+2}.
\end{eqnarray}
Hence, for a Poisson process, $\lambda_A/\lambda_B$ is not exactly
equal to $a/b$. Here the most probable value is taken as the
estimate of $\lambda_A/\lambda_B$. For the hardness ratio defined
as $HR'=\frac{\lambda_B-\lambda_A}{\lambda_B+\lambda_A}$, the
probability density function is,
\begin{eqnarray}
p(\frac{\lambda_B-\lambda_A}{\lambda_B+\lambda_A}=z|n_A=a, n_B=b)
&=& \frac{(1-z)^a(1+z)^b(a+b+1)!}{2^{(a+b+1)}a!b!}.
\end{eqnarray}
Again the most probable value $z_0$ is taken as the estimate of
$\frac{\lambda_B-\lambda_A}{\lambda_B+\lambda_A}$,
\begin{eqnarray}
{z_0}=(\frac{\lambda_B-\lambda_A}{\lambda_B+\lambda_A})_{\rm
peak}&=&\frac{b-a}{b+a}.
\end{eqnarray}
The method of error estimation is shown as the following equation
(the error interval denoted as [C, D]),
\begin{eqnarray}
\int_{\rm C}^{z_0}
p(\frac{\lambda_B-\lambda_A}{\lambda_B+\lambda_A}=z){\rm d}z &=&
90\%\times\int_{-1}^{z_0}
p(\frac{\lambda_B-\lambda_A}{\lambda_B+\lambda_A}=z){\rm d}z,
\end{eqnarray}
\begin{eqnarray} \int_{z_0}^{{\rm
D}}p(\frac{\lambda_B-\lambda_A}{\lambda_B+\lambda_A}=z){\rm d}z
&=& 90\%\times\int_{z_0}^{1}
p(\frac{\lambda_B-\lambda_A}{\lambda_B+\lambda_A}=z){\rm d}z.
\end{eqnarray}

\begin{figure}[tb]
\centerline{\epsfxsize 64mm \epsffile{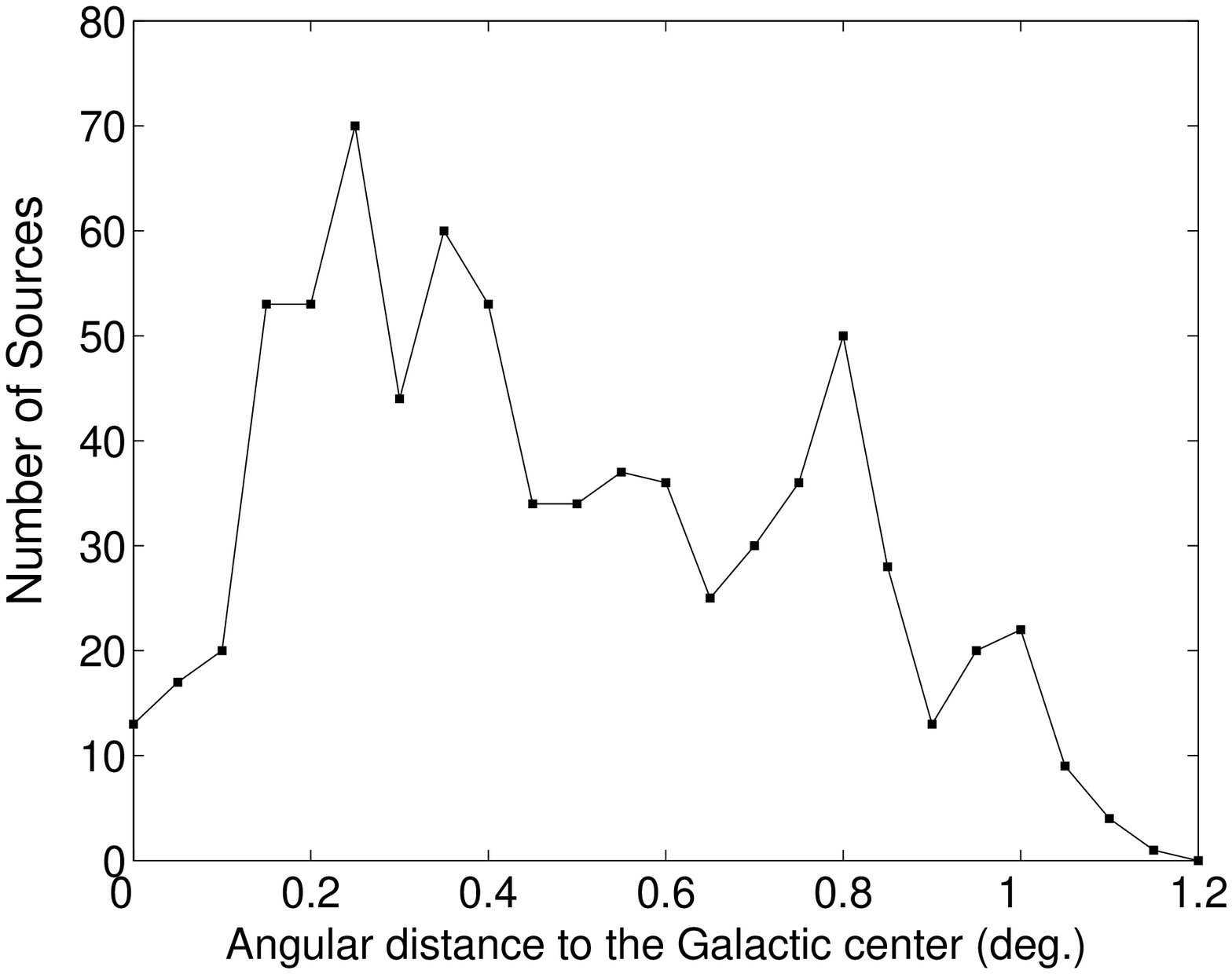}\ \ \ \ \epsfxsize
65mm \epsffile{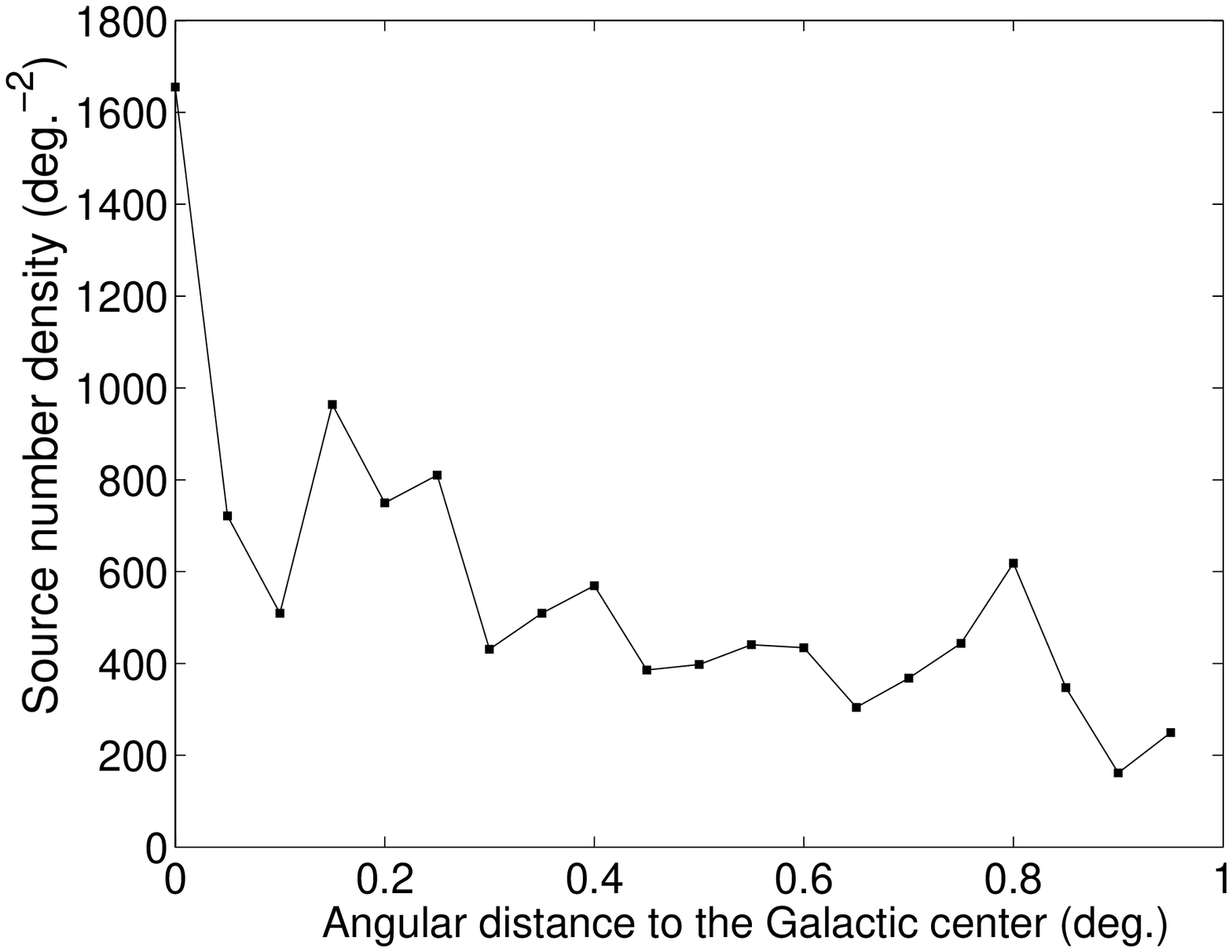}} \caption[]{Angular distribution of {\it
Chandra} GCS sources. Left panel: source count in each interval of
angular distance $[0^\circ, 0.05^\circ),\ [0.05^\circ,
0.10^\circ)\ ...\ [1.20^\circ, 1.25^\circ)$; Right panel: source
number density of each interval of angular distance $[0^\circ,
0.05^\circ),\ [0.05^\circ, 0.10^\circ)\ ...\ [0.95^\circ,
1.00^\circ)$.} \label{fig5}
\end{figure}

According to the relations in Fig. 3, the $N_H$ values and errors
for each source can be derived by linear interpolation. As can be
seen from Fig. 3, $ HR' $ and $HR2'$ estimates are more accurate
for $N_H < 2\times 10^{23}\ {\rm cm}^{-2}$ and $N_H
> 2\times 10^{23}\ {\rm cm}^{-2}$, respectively; thus the $N_H$ value for each
source are assigned accordingly. Because the curves in Fig. 3
cannot cover the whole range of $HR'$ and $HR2'$ as $[-1, 1]$, the
$N_H$ values of nearly 80\% ($\sim 600$) of the GCS sources are
acquired in this procedure, which are used in the following
investigation. The $N_H$ distributions derived from $HR'$ and
$HR2'$ are shown in Fig. 4.

\section{Spatial distribution of GCS sources}
\label{sect:spat}

\subsection{Angular distribution}
Fig. 5 shows the angular distribution of these sources. For
angular distance in the $[0.25^\circ, 0.30^\circ)$ interval, there
are the largest number of sources. However on the source number
density curve, the Galactic center is the densest region. A sharp
decrease exists in the immediate outer range. For the region with
angular distance $> 0.1^\circ$, the decrease of source number
density is slower. The total decrease is slower than the
exponential trend.

\subsection{Radial distribution --- Where are the sources?}

\begin{figure}[tb]
\centerline{\epsfxsize 75mm \epsffile{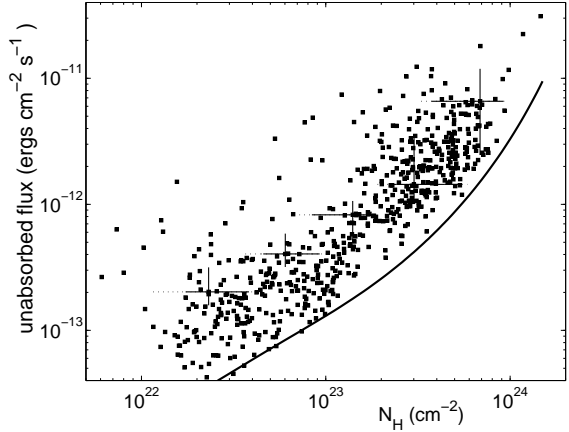}} \caption[]{The
correlation of unabsorbed flux of X-ray sources with HI column
density. The bold line shows the detection threshold of this
survey. The five crosses stand for the scales of the error bars of
$N_H$ and flux values. The extended dotted lines show the $N_H$
error intervals if the source photon indices $\Gamma$ vary in
$2\sim 3$ range.} \label{fig6}
\end{figure}

Our goal is to find out the three dimensional distribution of
these sources from the projected two dimensional image. It is
natural to consider that the sources are likely concentrated in
the Galactic center region since that region is known to be
crowded by various kinds of objects. Another possibility is that
the major part of the sources are distributed throughout the
Galactic disk. Here Monte Carlo simulation is used to study that
which model is consistent with the statistical properties of these
sources.

We start with the Fig. 6. The horizontal axis of this figure is
the HI column density of these sources, while the vertical axis is
the unabsorbed flux in 0.2-10.0 keV (all the fluxes hereafter are
in the same energy band), calculated from the $\Gamma=2.5$
power-law model, the corresponding HI column density and the count
rate of each source. The blank region in the bottom right corner
reflects the sensitivity of this survey. The criterion of the
source detection is based on Poisson probability (Wang 2004). The
probability of a count deviation above the background $C_{bg}$ can
be formulated as
\begin{eqnarray}
P=1-\sum_{n=0}^{C_{tl}-1}\frac{C_{bg}^n}{n!}e^{-C_{bg}},
\end{eqnarray}
where $C_{tl}$ and $C_{bg}$ are the total counts and background
counts for each source, respectively. If the probability $P$ is
less than a preset threshold $P_{th}$, the source detection could
be declared. Here in this {\it Chandra} GCS, the $P_{th}$ is
$10^{-5}$. In our case, this threshold could be approximated as
the $S/N>2$, where $S/N = {( C_{tl}-C_{bg})/\sqrt{C_{tl}}}$. This
point can be seen in the following simulations. The bold curve in
Fig. 6 shows the detection threshold in this approximation.

We design some simulated observations. We assume a set of sources with certain
luminosity function, and have them spread on the disk or concentrated at the Galactic
center. The parameters in the simulation are: 1) cumulative luminosity function:
power-law with $\Gamma=0.5$, which is similar to the discrete X-ray sources in nearby
spiral galaxies (e.g. Colbert {\it et al.} 2004; Tennant {\it et al.} 2001); 2) HI
column density distribution: For the case that these sources are concentrated in the
Galactic center, we simply take a uniform distribution in order to minimize the number
of simulation parameters; For the case that sources are spread throughout the Galactic
disk, the HI column density is assumed to be proportional to the radial distance,
taking the HI column density of the Galactic center as $1\times 10^{23}$ cm$^{-2}$
corresponding to 8.0 kpc distance (Baganoff {\it et al.} 2003); the farthest boundary
of the Milky Way to us is within 30.0 kpc; 3) background count distribution: the same
as the {\it Chandra} GCS observation, i.e., a Poisson distribution.

\begin{figure}[tb]
\centerline{\epsfxsize 145mm \epsfysize 60mm \epsffile{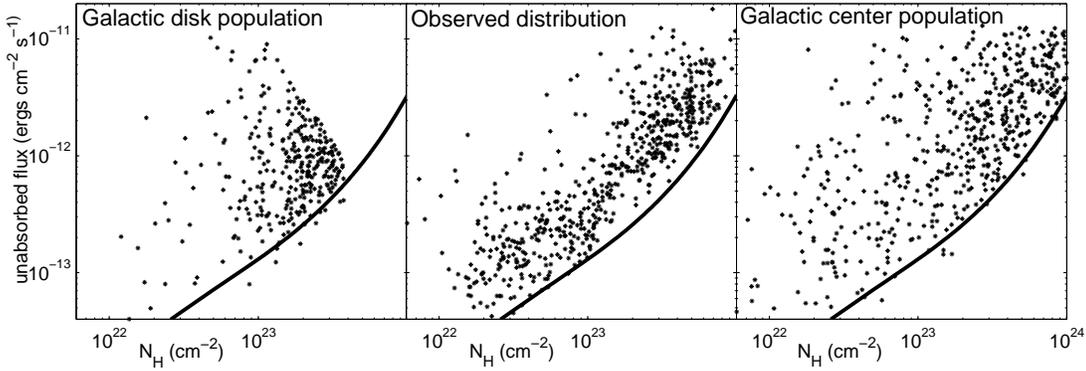}}
\caption[]{HI column density vs unabsorbed flux, where bold curves
in all three panels show the detection threshold of the real
survey. {\it Left panel:} simulation by assuming sources are
uniformly spread all over the Galactic disk. The value of HI
column density is proportional to its radial distance. {\it Middle
panel:} in real {\it Chandra} survey. {\it Right panel:}
simulation by assuming most sources concentrated in the Galactic
center. The value of HI column density is uniformly distributed.}
\label{fig7}
\end{figure}

Fig. 7 and Fig. 8 summarize the simulation results. $\chi^2$ test
is used to find out which simulation is closer to the observation.
For two discrete data sets, data in each set are divided into
several intervals. The number of intervals is $\nu$. $R_i$ and
$S_i$ are the numbers of data in the $i$th interval for each set,
respectively. Then the $\chi^2$ value
 is defined as the following equations (Press {\it et al.} 1992),

\begin{eqnarray}
\chi^2=\sum_i\frac{(\sqrt{S/R}R_i-\sqrt{R/S}S_i)^2}{R_i+S_i},
\end{eqnarray}
\begin{eqnarray}
R\equiv\sum_iR_i,\ \ \ \ \ \ \ \ \ \ S\equiv\sum_iS_i.
\end{eqnarray}
The degree of freedom for this $\chi^2$ distribution is $\nu$.

\begin{figure}[tb]
\centerline{\epsfxsize 75mm \epsffile{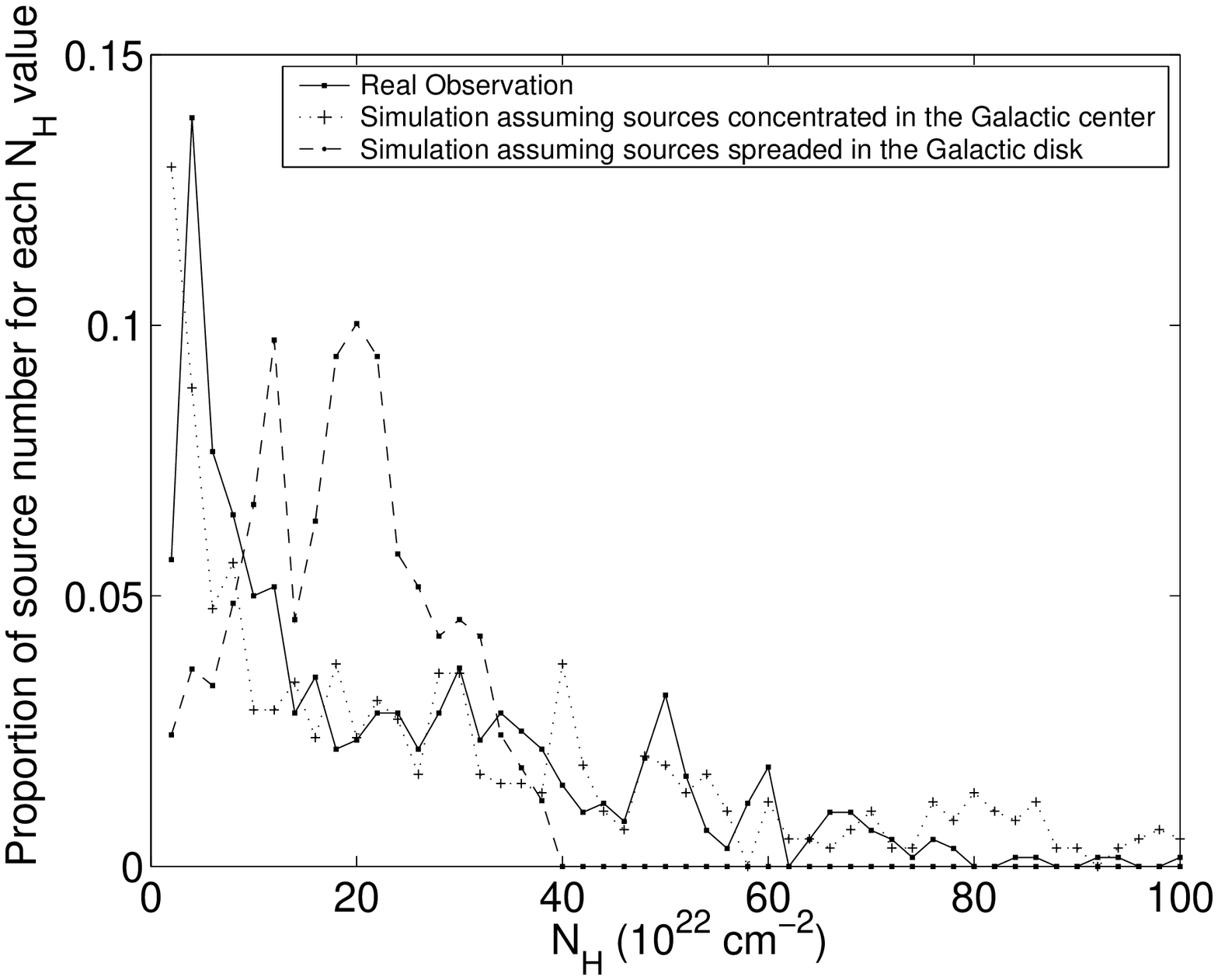}\ \ \epsfxsize 73mm
\epsffile{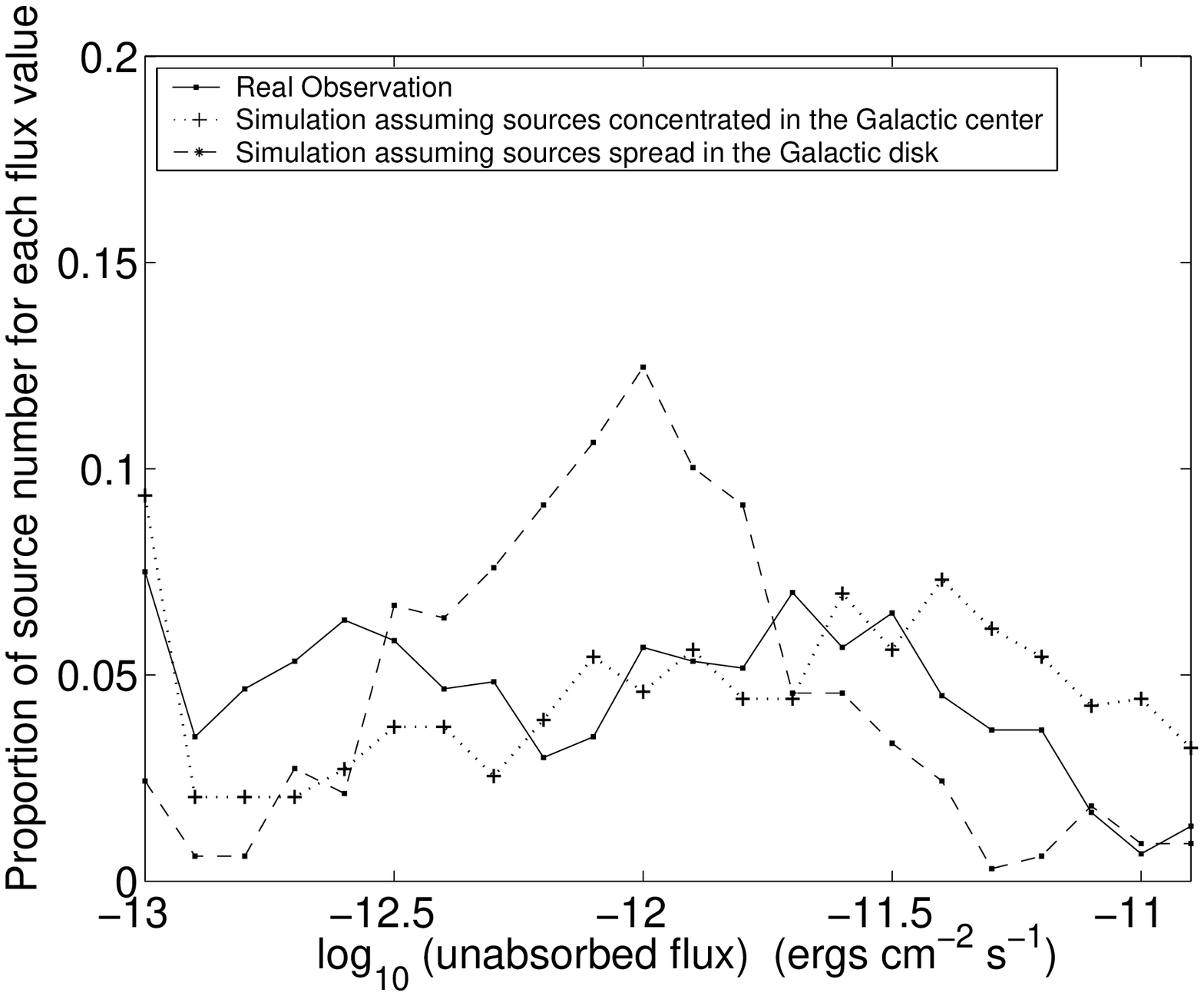}} \caption[]{{\it Left panel:} Comparison of $N_H$ value
distributions in real observation and simulations. The simulation in source
concentrating scenario is closer to the observation. The main difference shows in low
$N_H$ value interval. {\it Right Panel:} Comparison of unabsorbed flux value
distributions in real observation and simulations. Again the simulation in source
concentrating scenario is more likely. The first point of each line stands for sources
with unabsorbed flux $<10^{-13}$ ergs cm$^{-2}$ s$^{-1}$, while the last point stands
for sources with unabsorbed flux $>10^{-11}$ ergs cm$^{-2}$ s$^{-1}$.} \label{fig8}
\end{figure}

First, we examine the case that X-ray sources are spread out in
the disk. Fig. 8 shows that the simulated HI column density
distribution and the unabsorbed flux distribution in this case are
very different from those of the observation: $\chi^2/dof=199/40$
for the $N_H$ distribution, $\chi^2/dof=126/22$ for the unabsorbed
flux distribution. The main disagreement is that there are too
many sources with HI column density at $2\times10^{23}-4\times
10^{23} \ {\rm cm}^{-2}$ in this simulation than in observation.
The reason is that the source number density does not vary with
the Galactic latitude, i.e. the number density is uniform in the
field of view. However the density of the Galactic material is
known to decrease for high latitude region. An exponentially
decreasing factor $\exp{(-z/z_0)}$ is then used. For smaller $z_0$
value, the $\chi^2/dof$ value is found to decrease: even for
$z_0\sim 1$ pc, the simulated distribution of HI column density
cannot satisfactorily match the observation ($\chi^2/dof=111/40$).
Such small value of $z_0$ is inconsistent with our knowledge of
Galactic mass distribution: for various kinds of matter, $z_0\sim
70-400$ pc (Ferri\`{e}re 2001). The mass model of the Milky Way
(Bahcall \& Soneira 1980; Dehnen \& Binney 1998; Grimm, Gilfanov
\& Sunyaev 2002)is also tested and simulation cannot match the
observation either ($\chi^2/dof=294/40$ for the $N_H$
distribution). Furthermore, there are many differences between the
left panel and middle panel of Fig. 7. A void appears in the top
right part of the simulation figure, which is not present in the
observation. The reason for this void is that for the same
luminosity range, the flux of a farther source is smaller. There
is no reason to believe that farther sources are more luminous.
Therefore this difference cannot be accounted for easily. In fact,
we find that this kind of void exists whether or not we add the
exponentially decreasing factor. We thus conclude that the sources
in $Chandra$ GCS are not likely distributed throughout the
Galactic disk.

Then we examine the other case: Point-like sources are
concentrated in the Galactic center. The HI column density
distribution is assumed to be a uniform distribution. Comparisons
with the observed HI column density and flux distributions are:
$\chi^2/dof=104/40$ and $\chi^2/dof=83/22$, indicating significant
improvement over the case that these point sources are distributed
throughout the galactic disk. However even this case cannot match
the real observation completely. The middle and right panels of
Fig. 7 show the differences. One major difference is on the top
left part of the two panels. In observation, there is a void on
the top left part, i.e. less high flux sources with small HI
column density are detected than in simulation. This void cannot
be explained by selection effect, since selection effect only
makes sources under the threshold curve undetectable. Therefore,
from this point we infer that some kind of positive correlation
exists between the HI column density and the unabsorbed flux, i.e.
brighter sources have higher HI column density. Since these
sources are probably concentrated in the Galactic center, they
should have approximately the same distances. We therefore
conclude that a positive correlation exists between the X-ray
luminosities and HI column densities of the Galactic center
sources. The lack of high flux sources with low HI column density
values might also be due to the pileup effect for low $N_H$
sources or the cumulative luminosity function steeper than 0.5.
However, since there are very few sources (less than 1\%) which
have significant pileup effect, this kind of void is not likely
due to the pileup effect. 

\section{A potential interpretation for the nature of these sources}
\label{sect:nature}
As shown in the previous section, the luminosity of the X-ray
sources is positively correlated to the HI column density. One
possibility is that the X-ray luminosity comes from the
interaction between an isolated neutron star and interstellar
medium (ISM), mainly dense molecular clouds in the Galactic center
region.

Since stars in the Galactic center are very old, there should be
many dead isolated neutron stars. They may move into the molecular
clouds with speed $\sim 10^2$ km s$^{-1}$ (Arzoumanian {\it et
al.} 2002), and then accrete the ISM via standard Bondi accretion.
The statistical properties of molecular clouds in Galactic center
region (Miyazaki \& Tsuboi, 2000; Oka et al, 2001) are as the
following: 1) mass $M$: $1.3\times 10^4 $ --- $ 1.3\times 10^8 \
M_{\odot} $; 2) size $R$: $1.3 $ --- $ 50$ pc; 3) mass density
$\rho$: $10^{-22} $
--- $ 10^{-18}$ g cm$^{-3}$; 4) number density $n$: $10^2 $ --- $
10^6$ cm$^{-3}$. The sound speed of the molecular clouds ($\sim
1-10\ {\rm km\ s}^{-1}$) is much smaller than the velocity of the
moving neutron star. The accretion rate of standard Bondi
accretion should be calculated as
\begin{eqnarray}
\dot{M}\sim \frac{4\pi(GM_n)^2m_pn}{(V^2+C_S^2)^{3/2}}\sim 10^9 n
V_7^{-3}\ {\rm g}\ {\rm s}^{-1},
\end{eqnarray}
where $V_7$ is the velocity of neutron star taking unit as $10^7$
cm s$^{-1}$. Here we take $V_7=1$, neutron star mass $M_n=1.4\
M_\odot$, neutron star radius $R_n=10$ km, then the accretion
luminosity should be
\begin{eqnarray}
L_{acc}\sim \frac{GM_n\dot{M}}{R_n}\sim 10^{29}n\ {\rm ergs\
s}^{-1}.
\end{eqnarray}
Substitute the number density of molecular clouds into the above equation, the
luminosity should be $10^{31}-10^{35}$ ergs s$^{-1}$. For {\it Chandra} GCS sources
studied above, the unabsorbed fluxes stay in the range of $10^{-14}-10^{-11}$ ergs
cm$^{-2}$ s$^{-1}$. If most of these sources are at the Galactic center with a distance
of 8.0 kpc, the luminosities should be $10^{32}-10^{35}$ ergs s$^{-1}$, consistent with
our result. The values of column density from the center to the border of the dense
molecular clouds are $10^{22}-10^{24}$ cm$^{-2}$. Thus for the neutron stars in the
molecular clouds, the HI column density should be in this range. Since the molecular
clouds have different volumes, the HI column density is not strictly proportional to
the number density of a molecular cloud. However, a positive correlation does exist.
For sources with small HI column density, they are not likely to have high accreting
X-ray luminosity. Therefore in this scenario, the luminosity of X-ray sources is
positively correlated with the HI column density as in the observation. The top left
void in the middle panel of Fig. 7 can be explained in this scenario.

From the discussion above, we also argue that the absorption of X-ray is mainly caused
by the dense molecular clouds near the Galactic center, rather than by the Galactic
disk ISM along the line of sight to us. In fact, using the ISM model of the Milky Way
(Ferri\`{e}re 1998, Eqs.(6)(7)) the HI column density from the Galactic center to us is
calculated as only $\sim 10^{22}$ cm$^{-2}$, only a small fraction of the HI column
density of {\it Chandra} GCS sources. Our model can also explain why the HI column
density of sources in the Galactic center varies in a wide range. The HI column density
of foreground sources should be $< 10^{22}$ cm$^{-2}$. We can see in Fig. 6 that the
foreground sources are only a small part of the whole catalog. The percentage of
background extragalactic sources is also very small (Bandyopadhyay {\it et al.} 2005).
Thus the contribution from foreground and background sources can be ignored.

Treves et al. (2000) reviewed the theory and the observations of old isolated neutron
stars in the Milky Way. They estimated that there should be $10^8-10^9$ neutron stars
in our Galaxy, and most of them are dead. The Galactic center region is thus a good
place to observe old isolated neutron stars. Neutron stars in this region are so old
that the rotation frequency and magnetic field have decayed and the neutron stars could
be accretors, instead of ejectors and propellers. Even in ejector or propeller phase,
accretion could also occur (Zhang, Yu \& Zhang 1998): A small partion of materials
accretes onto the polar region of the neutron star through a quasi-spherical ADAF. This
scenario may occur in neutron star soft X-ray transients. For our model, the accretion
onto isolated neutron stars can also produce variability due to the instability of
magnetic field or the fine structure of ISM (Treves, private communication).

For those neutron stars still alive (i.e. radio pulsars), there
can also exist the positive correlation between their X-ray
luminosity and the ISM density. As these neutron stars move into
ISM with a speed much higher than the ISM sound speed, there
should be a bow shock. Romani, Cordes \& Yadigaroglu (1997)
discussed the Guitar nebula formed in this scenario. The shock
luminosity from the shocked ISM along the Guitar body is
proportional to the hydrogen number density. Recent research on
bow shocks also showed that for some region of the shock, the
X-ray luminosity is positively correlated with the ISM density
(Gaensler {\it et al.} 2004).

Further study on the nature of point-like X-ray sources needs
multiwavelength observation. Bandyopadhyay {\it et al.} (2005)
took a VLT infrared survey on 77 X-ray sources of the {\it
Chandra} GCS field. They detected candidate counterparts for 75\%
of the sources in their sample, and suggested that the X-ray
sources might be wind-fed accreting neutron star binaries.
However, their sample is only a small part of the sources. Laycock
et al. (2005) of ChaMPlane program excluded High Mass X-ray
Binaries (HXMBs) as the main contributor of X-ray sources in the
Galactic center. Other kinds of sources account for more than 90\%
of the X-ray sources. Furthermore, Belczynski \& Taam (2004)
argued that accreting neutron star systems are not likely to be
the main contributor to the faint X-ray sources in the Galactic
center, because neither wind-fed
 nor Roche lobe accreting neutron star systems can explain the amount
 or spectral properties of the sources in catalogs of
surveys in Wang, Gotthelf \& Lang (2002). Our model of the X-ray
sources, old isolated neutron stars with Bondi accretion, can
explain the faint soft X-ray source population in the Galactic
center.

\section{Summary}
\label{sect:summary}
We have performed statistical analysis to 600 of $\sim 800$
point-like sources detected in the Galactic center survey of {\it
Chandra X-ray Observatory} due to its excellent spatial resolution
and broad energy range. Fourteen bright sources detected are used
to fit jointly an absorbed power-law model, from which the
power-law photon index is determined to be $\sim$2.5. Assuming
that all other sources have the same power-law form, we use the
relation between hardness ratio and HI column density $N_H$ to
estimate the $N_H$ values for all sources. We study the spatial
distribution of these sources with Monte Carlo simulations and
infer that these sources are mainly concentrated in the Galactic
center region. The luminosities of X-ray sources are found to be
positively correlated with the HI column density. We propose a
model of old isolated neutron star with Bondi accretion as the
X-ray sources to explain the correlation. This model is consistent
with the present knowledge of dense molecular clouds in the
Galactic center. We also argue that the molecular clouds should be
the main contributor to X-ray absorption, rather than the ISM in
the Galactic disk. Further extensive multiwavelength surveys are
needed to study the nature of X-ray sources in the Galactic center
further.

\begin{acknowledgements}
We thank Dr. Q. D. Wang of University of Massachusetts for providing the GCS point-like
source list prior to its publication and his useful discussions and suggestions. We
also thank J.-L. Han, Y. Shen, F. Y. Bian, D. Lai, A. Treves, S. M. Tang and W. M.
Zhang for helpful discussions. X. Chen read the manuscript carefully and gave many
helpful suggestions, especially on English writing. This study is supported in part by
the Special Funds for Major State Basic Research Projects and by the National Natural
Science Foundation of China (project no.10521001).

\end{acknowledgements}

\label{lastpage}

\end{document}